\documentclass[10pt]{amsart}
\usepackage{amssymb,pstricks}
\usepackage{amsmath}
\usepackage{geometry}
\usepackage{amsfonts}

\topskip  \newtheorem{theorem}{Theorem}

\newtheorem{corollary}[theorem]{Corollary}

\newtheorem{example}[theorem]{Example}

\newtheorem{lemma}[theorem]{Lemma}

\newtheorem{proposition}[theorem]{Proposition}

\begin{document}
\title{Noncrossing Normal Ordering for Functions of Boson Operators}
\author{Toufik Mansour, Matthias Schork and Simone Severini}
\maketitle

\begin{center}
Department of Mathematics, University of Haifa, Haifa 31905, Israel\\[5pt]
Alexanderstrasse 76, 60489 Frankfurt, Germany\\[5pt]
Institute for Quantum Computing and Department of Combinatorics and Optimization,
University of Waterloo, N2L 3G1 Waterloo, Canada\\[5pt]

\textit{toufik@math.haifa.edu, mschork@member.ams.org, simoseve@gmail.com}
\end{center}


\section*{Abstract}

Normally ordered forms of functions of boson operators are important in many
contexts in particular concerning Quantum Field Theory and Quantum Optics.
Beginning with the seminal work of Katriel [\emph{Lett. Nuovo Cimento }%
\textbf{10(13)}:565--567, 1974], in the last few years, normally
ordered forms have been shown to have a rich combinatorial
structure, mainly in virtue of a link with the theory of partitions.
In this paper, we attempt to enrich this link. By considering linear
representations of noncrossing partitions, we define the notion of
\emph{noncrossing normal ordering}. Given the growing interest in
noncrossing partitions, because of their many unexpected connections
(like, for example, with free probability), noncrossing normal
ordering appears to be an intriguing notion. We explicitly give the
noncrossing normally ordered form of the functions $(a^{r}(a^{\dag
})^{s})^{n})$ and $(a^{r}+(a^{\dag })^{s})^{n}$, plus various
special cases. We are able to establish for the first time
bijections between noncrossing contractions of these functions,
$k$-ary trees and sets of lattice paths.

\textsc{PACS numbers}: 02.10.Ox

\section{Introduction}

Let $a$ and $a^{\dagger }$ be boson annihilation and creation operators,
satisfying the commutation relation $aa^{\dagger }-a^{\dagger }a=1$. The
normal ordering of an operator function $F(a,a^{\dagger })$ consists in
moving all creation operators to the left by applying the commutation
relation. The resulting expression is called the \emph{normally ordered form}
of $F(a,a^{\dagger })$ and it is denoted by $\mathcal{N}[F(a,a^{\dagger })]$%
. The expressions $F(a,a^{\dagger })$ and $\mathcal{N}[F(a,a^{\dagger })]$,
although generally different, represent the same function. The normally
ordered form is particularly useful in quantum optics \cite{ks} and Quantum
Field Theory \cite{b}. On the basis of Wick's theorem \cite{w}, one can
obtain $\mathcal{N}[F(a,a^{\dagger })]$ from $F(a,a^{\dagger })$ by means of
two operations: contraction and double-dot operation. A \emph{contraction}
consists of substituting $a=1$ and $a^{\dagger }=1$ in an expression
whenever $a$ precedes $a^{\dagger }$. An application of the \emph{double dot
operation} consists of deleting each occurrence of $1$ and then arranging
the expression so that $a^{\dagger }$ always precedes $a$. For example, $%
:a^{k}(a^{\dagger })^{n}:$ $=(a^{\dagger })^{n}a^{k}$. Among all possible
contractions, we also include the \emph{null contraction}, that is the
contraction leaving the expression invariant. Specifically,
\begin{equation}\label{normal}
F(a,a^{\dagger })\equiv \mathcal{N}[F(a,a^{\dagger })]=\sum :\{\text{all
possible contractions}\}:.
\end{equation}%
For example, if $F(a,a^{\dagger })=aa^{\dagger }aaa^{\dagger }aa$ then $%
\mathcal{N}[F(a,a^{\dagger })]=(a^{\dagger })^{2}a^{5}+4a^{\dagger
}a^{4}+2a^{3}$.

The combinatorics of normally ordered forms has been studied in a number of
papers (\cite{thesis} is a survey). For example, several authors established
connections between Stirling, Bell numbers and normally ordered forms (see~%
\cite{Kat02, v} and references therein). In fact, it is nowadays well-known
that $\mathcal{N}[\left( a^{\dagger }a\right)
^{n}]=\sum_{k=1}^{n}S(n,k)(a^{\dagger })^{k}a^{k}$, where the integers $%
S(n,k)$ are the so called Stirling numbers of second kind (see, \emph{e.g.}, \cite[Seq.
A008277]{Int}), satisfying the recurrence relation $%
S(n+1,k)=kS(n,k)+S(n,k-1) $, with $S(n,0)=\delta _{n,0}$ and $S(n,k)=0$ for $%
k>n$. Generally, it can be difficult to obtain $\mathcal{N}[F(a,a^{\dagger
})]$ when $F(a,a^{\dagger })$ is a polynomial of high order or an infinite
series \cite{gl}. A considerable amount of recent work has been produced in
this direction \cite{bd, bp, bh, bs}. Among the results, are explicit
formulas for many examples of operators depending on $(a^{\dag })^{k}a^{n}$,
characterized by integer powers $k$ and $n$, or depending on $q(a^{\dag
})a+v(a^{\dag })$, with arbitrary functions $q$ and $v$.

The picture of this scenario has two faces:\ techniques from combinatorics
are fruitfully applied to obtain normally ordered forms with immediate use
in physics (\emph{e.g.} the construction of generalized coherent states \cite%
{bps}); the physical machinery behind this context helps to unveil and
describe combinatorial properties (\emph{e.g.} a theory of the Stirling and
Bell polynomials can be formulated in terms of the algebraic and Fock space
properties of the boson operators \cite{Kat02}).

The purpose of the present paper is to introduce and study the notion of
noncrossing normally ordered form, an \emph{extremal} structure with a
well-defined combinatorial interpretation. Before entering this subject, it
is useful to describe two ways to represent contractions: a graphical
representation of contractions, which we call \emph{linear representation};
a representation as words which is called \emph{canonical sequential form}
(see \cite{Kl1}). Let $\pi _{1}\pi _{2}\ldots \pi _{n}$ be a word whose
letters are boson operators. Since our discussion is focused on the
combinatorial properties of boson functions, we consider $a$ and $a^{\dagger
}$ as letters, disregarding these as operators.

\begin{itemize}
\item \underline{Linear representation:} We draw $n$ vertices, say $%
1,2,\ldots ,n$, on a horizontal line, such that the point $i$ corresponds to
the letter $\pi _{i}$; we represent each $a$ by a white vertex and each $%
a^{\dagger }$ by a black vertex; each black vertex $i$, incident with an
edge, is connected by an undirected edge $(i,j)$ to a white vertex $j$.
Importantly, the edges are drawn above the points. We call this graphical
representation the \emph{linear representation} of the contraction or, with
an abuse of the terminology, just \emph{contraction}. The linear
representations of all contractions of the word $aaa^{\dagger }a^{\dagger }$
are illustrated in Figure~\ref{flat}.
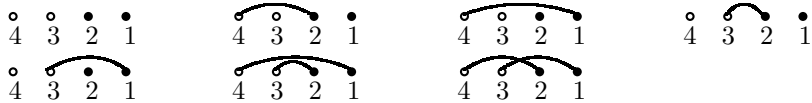
\begin{figure}[th]
\begin{center}
\begin{pspicture}(0,0)(10,.2)
\setlength{\unitlength}{3mm} \linethickness{0.3pt}
\multips(0,0)(3,0){4}{\pscircle(0,0){.2}\pscircle(.5,0){.2}\pscircle*(1,0){.2}\pscircle*(1.5,0){.2}}
\linethickness{0.8pt}\qbezier(10,.1)(11.6,1)(13.2,.1)
\qbezier(20,.1)(22.5,1)(25,.1)\qbezier(31.6,.1)(32.4,1)(33.2,.1)
\put(-.2,-1.2){$4$}\put(1.5,-1.2){$3$}\put(3.2,-1.2){$2$}\put(4.9,-1.2){$1$}
\put(9.7,-1.2){$4$}\put(11.4,-1.2){$3$}\put(13.1,-1.2){$2$}\put(14.8,-1.2){$1$}
\put(19.7,-1.2){$4$}\put(21.4,-1.2){$3$}\put(23.1,-1.2){$2$}\put(24.8,-1.2){$1$}
\put(29.7,-1.2){$4$}\put(31.4,-1.2){$3$}\put(33.1,-1.2){$2$}\put(34.8,-1.2){$1$}
\end{pspicture}
\par
\begin{pspicture}(0,0)(10,.7)
\setlength{\unitlength}{3mm} \linethickness{0.3pt}
\multips(0,0)(3,0){3}{\pscircle(0,0){.2}\pscircle(.5,0){.2}\pscircle*(1,0){.2}\pscircle*(1.5,0){.2}}
\linethickness{0.8pt}\qbezier(1.5,.1)(3.5,1.2)(5,.1)
\qbezier(10,.1)(12.5,1.2)(15,.1)\qbezier(11.6,.1)(12.4,.8)(13.2,.1)
\qbezier(20,.1)(21.65,1.2)(23.3,.1)\qbezier(21.6,.1)(23.25,1.2)(24.9,.1)
\put(-.2,-1.2){$4$}\put(1.5,-1.2){$3$}\put(3.2,-1.2){$2$}\put(4.9,-1.2){$1$}
\put(9.7,-1.2){$4$}\put(11.4,-1.2){$3$}\put(13.1,-1.2){$2$}\put(14.8,-1.2){$1$}
\put(19.7,-1.2){$4$}\put(21.4,-1.2){$3$}\put(23.1,-1.2){$2$}\put(24.8,-1.2){$1$}
\end{pspicture}
\end{center}
\caption{The linear representation of the contractions of the word $%
aaa^{\dagger }a^{\dagger }$}
\label{flat}
\end{figure}

\item \underline{Canonical sequential form:} A contraction $\pi$ is
represented by a sequence $a_{1}a_{2}\ldots a_{n}$ on the set $\{1,2,\ldots
,n,1^{\prime},2^{\prime},\ldots,n^{\prime}\}$. In order to construct the
sequence $a_{1}a_{2}\ldots a_{n}$, we need to read the contraction $\pi$
from right to left.

\begin{itemize}
\item If $\pi_{j}$ is a white (resp. black) vertex of degree $0$ (that is,
incident with no edges) we then replace it with $i^{\prime}$ (resp. $i$); $i$
is the smallest number not appearing in the sequence.

\item If $\pi_{j}$ is a black vertex of degree $1$ we then replace it with $%
i $, where $i$ is the smallest number not appearing in the sequence.

\item If $\pi_{j}$ is a white vertex of degree $1$ we then replace it with $%
i $, where $i$ is associated to the black vertex connected to $\pi_{j}$.
\end{itemize}

For example, the contractions in Figure~\ref{flat} can be represented by $%
123^{\prime }4^{\prime }$, $123^{\prime }2$, $123^{\prime }1$, $1223^{\prime
}$, $1213^{\prime }$, $1221$, and $1212$. Such a representation is called
the \emph{canonical sequential form}~\cite{Kl1}.
\end{itemize}

We denote contractions also by enumerating the edges. For example, the contractions in Figure~\ref{flat} are given by $\{\emptyset, (42),(41),(32),(31),(41)(32),(42)(31)\}$. Let $e=(i,j)$ and $e^{\prime }=(p,q)$ be two edges of a contraction $\pi $.
We say that $e$ \emph{crosses} $e^{\prime }$ if they intersect with each
other, or, in other words, if $i<p<j<q$ or $p<i<q<j$. If this is the case, $e$ and $%
e^{\prime }$ are said to be a \emph{crossing} of the contraction; otherwise,
$e$ and $e^{\prime }$ are said to be a \emph{noncrossing}. For example, from the contractions in Figure~\ref{flat} only $(42)(31)$ is a crossing.

Using the linear representation form of contractions, the normally ordered form of an expression $F(a,a^{\dag })$ can be defined as follows. Given a
contraction $\pi $ associated with its linear representation form, let $%
freeb(\pi )$ (resp. $freew(\pi )$) to be number black (resp. white) vertices
of degree zero in $\pi $. The normally ordered form of $F(a,a^{\dag })$ is
given by
\begin{equation*}
\sum_{\pi \mbox{ is a contraction of
}F(a,a^{\dag })}(a^{\dag })^{freeb(\pi )}a^{freew(\pi )}.
\end{equation*}%
A \textquotedblleft $p$-analogue\textquotedblright\ of the normally ordered
form can be defined as follows. Let us introduce an operator $P_{p}$ acting
on each contraction $\pi $ of an expression $F=F(a,a^{\dag })$:
\begin{equation*}
P_{p}(\pi ):=p^{cross(\pi )}{(a^{\dag })^{freeb(\pi )}}a^{freew(\pi )},
\end{equation*}%
where $cross(\pi )$ counts the number
of crossing edges in $\pi $. We extend $P_{p}$ to a linear operator by
\begin{equation*}
P_{p}(F):=\sum_{\pi }P_{p}(\pi ),
\end{equation*}%
where the sum is taken over all contractions of the expression $F$. Note
that the operator $P_{p}$ is a $p$-analogue\ of the standard double dot
operation. Namely, for a given expression $F$, $P_{1}(F)$ is exactly the
normally ordered form of $F$.

The \emph{number-states} $\mid m\rangle $ are the states that satisfy $%
a^{\dag }a\mid m\rangle =m\mid m\rangle $, where $\langle m\mid m\rangle =1$%
. The \emph{coherent states} $\mid \gamma \rangle $ are the states that
satisfy $a\mid \gamma \rangle =\gamma \mid \gamma \rangle $, where $\langle
\gamma \mid \gamma \rangle =1$. Number-states and coherent states are
important within the boson Fock space (see \cite{b}). These are related by
the well-known expression $\mid \gamma \rangle =e^{-|\gamma
|^{2}/2}\sum_{m\geq 0}\frac{\gamma ^{m}}{\sqrt{m!}}\mid m\rangle $. Many
authors (see~\cite{bd,bp,bps,bh,bs,Kat74,Kat83,Kat00,Kat02}) dealt with
explicit formulas for the expectation values $\langle \gamma \mid
F(a,a^{\dag })\mid \gamma \rangle $. For example, Katriel~\cite{Kat02}
showed that $\langle \gamma \mid (a^{\dag }a)^{n}\mid \gamma \rangle
=\sum_{j=0}^{n}S(n,j)|\gamma |^{2j}$, where $S(n,j)$ are the Stirling numbers of the second kind. This was done by
proving that the normally ordered form of $(a^{\dag }a)^{n}$ is given by $%
\sum_{j=0}^{n}S(n,j)(a^{\dag })^{j}a^{j}$. By applying our operator $P_{p}$ to the
expression $(a^{\dag }a)^{n}$, one can write
\begin{equation*}
P_{p}((a^{\dag }a)^{n})=\sum_{j=0}^{n}f_{n,j}(p)(a^{\dag })^{j}a^{j},
\end{equation*}%
where $f_{n,j}(p)$ is a polynomial in $p$ (note that $f_{n,j}(1)=S(n,j)$). In this paper, we study the combinatorial structure of $%
P_{0}(F)$. In particular, with Theorem 3, we show that
\begin{equation*}
P_{0}\biggl((aa^{\dag })^{n}\biggr)=\sum\limits_{j=0}^{n}\frac{1}{n+1}\binom{%
n+1}{j+1}\binom{n+1}{j}(a^{\dag })^{n-j}a^{n-j}.
\end{equation*}
If the commutation relations hold for the operators then one has $aa^{\dag}=a^{\dag}a+1$ and, therefore, a close relation between $(aa^{\dag})^n$ and $(a^{\dag}a)^n$.

A \emph{noncrossing contraction} is a contraction whose edges are all
noncrossing. Figure~\ref{flat} gives all contractions of the word $aaa^{\dag
}a^{\dag }$: these consist of $1$ crossing and $6$ noncrossing contractions
(one of those is the null contraction). With this terminology, we are now
ready to give the definition of noncrossing normal ordering.\bigskip

\noindent\textbf{Definition.}\emph{\ Let }$F(a,a^{\dag})=\sum_{w\in A}w$%
\emph{\ be any expression composed by elements of }$A$\emph{, where }$A$%
\emph{\ is any finite set of finite words on the alphabet }$\{a,a^{\dag}\}$%
\emph{. The }noncrossing normally ordered form\emph{\ of }$F(a,a^{\dag})$%
\emph{\ is denoted and defined by}%
\begin{equation}\label{nonormal}
\mathcal{NC}(F(a,a^{\dag}))=\sum:\{\text{all possible noncrossing
contractions}\}:
\end{equation}

\bigskip

It is important to note that
$P_{0}(F(a,a^{\dag}))=\mathcal{NC}(F(a,a^{\dag }))$. The central
concept of the paper is then a \textquotedblleft noncrossing
combinatorial structure\textquotedblright. A number of aspects of
noncrossing combinatorial structures have been studied in recent
years. In particular, noncrossing partitions have recently turned
out to be a flourishing subject, given the links with other concepts
like free probability, parking functions, and braid groups (see
\cite{mc,sim}). Specifically, in enumerative combinatorics, Klazar
linked noncrossing partitions to the well-known Catalan numbers (or
the lattice Dyck paths) \cite{Kl1, Kl2}, while other authors have
related noncrossing
partitions to RNA secondary structure and Motzkin paths (%
\emph{e.g.}, see \cite{bb} and references therein). Here, we apply the
notion of linear representation of noncrossing partitions to contractions of
normally ordered forms. In this way, the obtained \emph{noncrossing}
normally ordered forms can be related to a number of different combinatorial
objects. Namely, we study $\mathcal{NC}((a^{r}(a^{\dag})^{s})^{n})$ (in
Section 3), $\mathcal{NC}((a^{r}+(a^{\dag })^{s})^{n})$ (in Section 4)
and some special cases. In Section 5, we establish bijections between sets
of noncrossing contractions of special cases of these functions (for
example, $(a+(a^{\dagger})^{2})^{n}$, $(a^{r}a^{\dagger})^{n}$ and $%
(a(a^{\dag})^{r})^{n}$) sets of trees and sets of lattice paths (for
example, $k$-ary trees and 2-Motzkin paths). Section 6, the last section of
the paper, is devoted to draw some directions for further analysis. Physical
interpretations of the noncrossing normal ordering remain as \emph{desiderata%
}.

\section{Some preliminary observations}\label{Prelim}
Let us denote for arbitrary $p$ the corresponding normal ordering by
\[
\mathcal{N}_p(F(a,a^{\dag}))\equiv P_p(F(a,a^{\dag}))
\]
such that $\mathcal{N}_1\equiv \mathcal{N}$ (the conventional normal ordering) and $\mathcal{N}_0\equiv \mathcal{NC}$ (the noncrossing normal ordering). If we denote the associated coefficients by $C_{F;k,l}(p)$, i.e.,
\[
\mathcal{N}_p(F(a,a^{\dag}))=\sum_{k,l}C_{F;k,l}(p)(a^{\dag})^ka^l,
\]
then it is clear that one has for any $F$ and all $k,l$ the following inequality
\[
0\leq C_{F;k,l}(0) \leq C_{F;k,l}(1).
\]
For example, if $F(a,a^{\dag})=(a^{\dag}a)^n$ then one has $C_{(a^{\dag}a)^n;k,l}(1)=S(n,k)\delta_{kl}$.

\begin{example} Let $F(a,a^{\dag})=aaa^{\dag}a^{\dag}$ be the example of Figure~\ref{flat}. Using the commutation relations or (\ref{normal}) one may show that $\mathcal{N}_1(aaa^{\dag}a^{\dag})=(a^{\dag})^2a^2+4a^{\dag}a+2$. Since there is exactly one crossing contraction of degree two, namely $(42)(31)$, one has $\mathcal{N}_0(aaa^{\dag}a^{\dag})=(a^{\dag})^2a^2+4a^{\dag}a+1$.
\end{example}
Generalizing the observation of the preceding example, it is clear that $\mathcal{N}_0(F(a,a^{\dag}))\neq \mathcal{N}_1(F(a,a^{\dag}))$ if and only if $F(a,a^{\dag})$ is a polynomial of degree at least two in $a$ and $a^{\dag}$. In particular, for lower degrees the two expressions coincide. This has the consequence that $\mathcal{N}_0(aa^{\dag})=a^{\dag}a+1$. Recall that in the conventional case one also has $\mathcal{N}_1(aa^{\dag})=a^{\dag}a+1$ directly from (\ref{normal}). On the other hand, one can use the commutation relation to find the same result, i.e., $\mathcal{N}_1(aa^{\dag})=\mathcal{N}_1(a^{\dag}a+1)=a^{\dag}a+1$. Let us try to reproduce the noncrossing statistics according to (\ref{nonormal}) by modifying the commutation relations and using the usual normal ordering process. Thus, we consider operators $b,b^{\dag}$ satisfying
\begin{equation}\label{commu}
bb^{\dag}-\kappa b^{\dag}b=\lambda_1b^{\dag}+\lambda_2b+\lambda_3
\end{equation}
where $\kappa$ and $\lambda_i$ are some parameters (here we have assumed that the right-hand side has lower degree than the left-hand side). Using this commutator, it follows that
\[
\mathcal{N}_0(bb^{\dag})\stackrel{(\ref{commu})}{=}\mathcal{N}_0(\kappa b^{\dag}b+\lambda_1b^{\dag}+\lambda_2b+\lambda_3)=\kappa b^{\dag}b+\lambda_1b^{\dag}+\lambda_2b+\lambda_3
\]
where we have used in the second equation the fact that $\mathcal{N}_0$ is linear and that all summands are already normally ordered. However, if this has to be the result obtained from the definition (\ref{nonormal}) then one has neccessarily $\kappa=1=\lambda_3$ as well as $\lambda_1=\lambda_2=0$, reproducing for $b,b^{\dag}$ the conventional commutation relations. However, the explicit examples show that the operators $b,b^{\dag}$ cannot satisfy the conventional commutation relations and the noncrossing normal ordering (\ref{nonormal}) simultanously! Thus, it is not clear whether one should speak of the operators $a,a^{\dag}$ - whose words $F(a,a^{\dag})$ are brought into noncrossing normally ordered form using (\ref{nonormal}) - as ``bosonic'' operators anymore. It would be very interesting to find the algebraic relations which the operators have to satisfy such that normal ordering using these algebraic relations is equivalent to the definition (\ref{nonormal}). A discussion of a general approach to generalizations of normal orderings using nearly arbitrary weights can be found in \cite{tsm}.

Let us mention that very closely related situations have been considered in the context of $q$-Fock spaces and $q$-Gaussian processes when the limit $q\rightarrow 0$ is considered, see, e.g., \cite{ansh,speicher,effros} and the references given therein. Here the corresponding annihilation and creation operators satisfy the $q$-deformed commutation relations. In the limit $q\rightarrow 0$ one considers full Fock space and there exists an intimate link between normal ordered representations and noncrossing partitions.

\section{Noncrossing normal ordering of $(a^{r}(a^{\dagger})^{s})^{n}$}

Let us denote by $\mathcal{V}_{rs}(n)$ the set of all the linear
representations of the noncrossing contractions of $(a^{r}(a^{%
\dagger})^{s})^{n}$. For each linear representation $\pi\in\mathcal{V}%
_{rs}(n)$, define $e(\pi)$ to be the number of edges in $\pi$. Let $%
B_{r,s}(x,y)$ be the generating function for the number of linear
representations $\pi \in\mathcal{V}_{rs}(n)$ with exactly $m$ edges, that
is,
\begin{equation*}
B_{r,s}(x,y)=\sum_{n\geq0}\sum_{\pi\in\mathcal{V}_{rs}(n)}x^{n}y^{e(\pi)}.
\end{equation*}
The noncrossing normally ordered form of $(a^{r}(a^{\dag})^{s})^{n}$ is
given by
\begin{equation*}
\mathcal{NC}\biggl((a^{r}(a^{\dag})^{s})^{n}\biggr)=\sum_{j\geq0}(%
\mbox{the
coefficient of }x^{n}y^{j}\mbox{ in
}B_{r,s}(x,y))(a^{\dag})^{sn-j}a^{rn-j}.
\end{equation*}
So, to find the noncrossing normally ordered form of $(a^{r}(a^{%
\dag})^{s})^{n}$, it is enough to find an explicit formula for the
generating function $B_{r,s}(x,y)$. In this section, we present a nonlinear
system of equations whose solution gives an explicit formula for $%
B_{r,s}(x,y)$. We begin by writing
\begin{equation}
B_{r,s}(x,y)=1+B_{r,s}(x,y;s),  \label{eqa1}
\end{equation}
where $B_{r,s}(x,y;t)$ is the generating function for all the linear
representations $\pi=\pi_{1}\ldots\pi_{n}\in\mathcal{V}_{rs}(n)$, such that
the canonical sequential form of $\pi$ starts with $12\ldots t$. \ With the
following lemma, we give a recurrence relation for the sequence $%
B_{r,s}(x,y;t)$.

\begin{lemma}
\label{lemaa1} Let $z^{\prime}=x^{\frac{1}{r+s}}$. For all $t=1,2,\ldots,s$,
\begin{equation*}
B_{r,s}(x,y;t)=z^{\prime}B_{r,s}(x,y;t-1)+B_{r,s}(x,y)\sum_{j=1}^{r}{%
z^{\prime}}^{r-j}B_{r,s}(x,y;j,t),
\end{equation*}
with the initial condition $B_{r,s}(x,y;0)={z^{\prime}}^{r}B_{r,s}(x,y)$,
where $B_{r,s}(x,y;a,b)$ is the generating function for the number of linear
representations $\pi\in\mathcal{V}_{rs}(n)$ such that $\pi$ starts with $b$
black vertices and end with $a$ white vertices and there is an edge between
the first black vertex and last white vertex.
\end{lemma}

\begin{proof}
Let $\pi$ be any linear representations in $\mathcal{V}_{rs}(n)$ such that $%
\pi$ starts with $t$ black vertices. Let us write an equation for $%
B_{r,s}(x,y;t)$. The first black vertex has degree either zero or one. The
contribution of the first case is%
\begin{equation*}
z^{\prime}B_{r,s}(x,y;t-1).
\end{equation*}

Now, let us consider the second case, that is the first black vertex having
degree one. The contraction $\pi$ can be written as $\pi=\pi^{\prime
}(p+1)^{\prime}\ldots q^{\prime}\pi^{\prime\prime}$, where $\pi^{\prime}$
starts with $t$ black vertices, ends with $j$ white vertices and there is an
edge between the first black vertex of $\pi^{\prime}$ and the last white
vertex of $\pi^{\prime}$. Moreover, $\pi^{\prime}$ is followed by $q-p$
white vertices such that $q-p+j=r$, see the following figure:

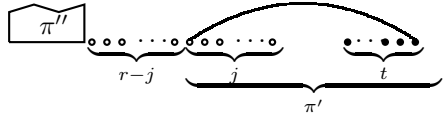
\begin{figure}[ht]
\begin{center}
\begin{pspicture}(0,-.5)(5.5,.5)
\psline(0,0)(1,0)(1,.5)(.66,.4)(.33,.5)(0,.4)(0,0)
\put(0.4,.1){$\pi''$}
\multips(0,0)(0.2,0){3}{\pscircle(1.1,0){0.05}}\put(1.7,0){$\ldots$}
\multips(1.1,0)(0.2,0){4}{\pscircle(1.1,0){0.05}}\put(3,0){$\ldots$}\pscircle(3.5,0){0.05}
\pscircle*(4.5,0){0.05}\put(4.6,0){$\ldots$}\multips(3.9,0)(0.2,0){3}{\pscircle*(1.1,0){0.05}}
\linethickness{0.7pt}\qbezier(5.4,0.03)(3.9,1)(2.4,.03)
\put(4.45,-.03){$\underbrace{\qquad\quad}_{ t}$}
\put(1.05,-.03){$\underbrace{\qquad\quad\,\,\,\,}_{ r-j}$}
\put(2.35,-.03){$\underbrace{\qquad\quad\,\,\,\,}_{j}$}
\put(2.35,-.45){$\underbrace{\qquad\quad\qquad\quad\qquad\quad\,\,\,\,}_{\pi'}$}
\end{pspicture}
\end{center}
\par
\label{figlemaa1}
\end{figure}

Since we are interested in the noncrossing contractions, we may observe
there are no edges from $\pi ^{\prime }(p+1)^{\prime }\ldots q^{\prime }$ to
$\pi ^{\prime \prime }$. Therefore, the contribution of the second case
gives
\begin{equation*}
B_{r,s}(x,y)\sum_{j=1}^{r}{z^{\prime }}^{r-j}B_{r,s}(x,y;j,t).
\end{equation*}

Considering these disjoint cases together we arrive to our recurrence
relation. Since there are no edges from white vertices to other vertices,
the generating function is $B_{r,s}(x,y;0)={z^{\prime}}^{r}B_{r,s}(x,y)$, as
required by the statement of the lemma.
\end{proof}

As we have seen in Lemma~\ref{lemaa1}, to find a formula for the generating
function $B_{r,s}(x,y)$, we need to have a recurrence relation for the
generating functions $B_{r,s}(x,y;p,q)$. This is done as follows.

\begin{lemma}
\label{lemaa2} Let $z^{\prime}=x^{\frac{1}{r+s}}$, $p=2,3,\ldots,r$, and $%
q=2,3,\ldots,s$. Then the following hold:

\textrm{(i)} $B_{r,s}(x,y;1,1)=y{z^{\prime}}^{2}+xyz^{2}B_{r,s}(x,y)$.

\textrm{(ii)} $B_{r,s}(x,y;p,1)=z^{\prime}B_{r,s}(x,y;p-1,1)+y{z^{\prime}}%
^{r+2}B_{r,s}(x,y)\sum\limits_{j=1}^{s}{z^{\prime}}^{s-j}B_{r,s}(x,y;p-1,j)$.

\textrm{(iii)} $B_{r,s}(x,y;1,q)=z^{\prime}B_{r,s}(x,y;1,q-1)+y{z^{\prime}}%
^{s+2}B_{r,s}(x,y)\sum\limits_{j=1}^{r}{z^{\prime}}^{r-j}B_{r,s}(x,y;j,q-1)$.

\textrm{(iv)} $%
\begin{array}{ll}
B_{r,s}(x,y;p,q)=z^{\prime}B_{r,s}(x,y;p-1,q) & +y{z^{\prime}}^{2}
\sum\limits_{j=1}^{q-1}{z^{\prime}}^{q-1-j}B_{r,s}(x,y;p-1,j) \\
& +y{z^{\prime}}^{2}B_{r,s}(x,y;q-1)\sum\limits_{j=1}^{s}{z^{\prime}}%
^{s-j}B_{r,s}(x,y;p-1,j).%
\end{array}
$
\end{lemma}

\begin{proof}
Let $\pi$ be any linear representations in $\mathcal{V}_{rs}(n)$ such that
the last $p$ vertices of $\pi$ are white, the first $q$ vertices of $\pi$
are black and there is an edge between the first black vertex and the last
white vertex. The generating function for the number of such linear
representations $\pi$ is given by $B_{r,s}(x,y;p,q)$. Now, let us write an
equation for $B_{r,s}(x,y;p,q)$ for each of the following four cases:

\begin{itemize}
\item If $p=q=1$ then there are two possibilities for the linear
representation $\pi$: $\pi=11$ or $\pi=12^{\prime}3^{\prime}\ldots r^{\prime
}\pi^{\prime}(d+1)(d+2)\ldots(d+s)1$. The first contribution gives ${%
z^{\prime}}^{2}y$ and the second contribution gives $y{z^{\prime}}%
^{r+s+2}B_{r,s}(x,y)=yx{z^{\prime}}^{2}B_{r,s}(x,y)$. Putting together the
two disjoint cases we obtain \emph{(i)}.

\item If $p\geq2$ and $q=1$ then the degree of the vertex $v$, the one
before the last white vertex (which is also a white vertex), is either zero or
one. The first contribution gives $z^{\prime}B_{r,s}(x,y;p-1,1)$. In the
second case, there exists a black vertex connected to $v$. Then $\pi$ can be
represented as

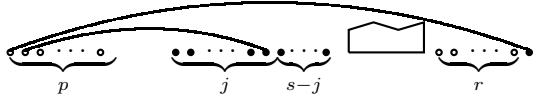
\begin{figure}[ht]
\begin{center}
\begin{pspicture}(0,-.1)(7,1)
\multips(0,0)(0.2,0){3}{\pscircle(0,0){0.05}}\put(.6,0){$\ldots$}\pscircle(1.2,0){0.05}
\multips(2.2,0)(0.2,0){2}{\pscircle*(0,0){0.05}}\put(2.6,0){$\ldots$}\pscircle*(3.2,0){0.05}
\multips(3.4,0)(0.2,0){2}{\pscircle*(0,0){0.05}}\put(3.7,0){$\ldots$}\pscircle*(4.2,0){0.05}
\multips(5.7,0)(0.2,0){2}{\pscircle(0,0){0.05}}\put(6.1,0){$\ldots$}\pscircle(6.7,0){0.05}\pscircle*(6.9,0){0.05}
\multips(4.5,0)(0,0){1}{\psline(0,0)(1,0)(1,.4)(.66,.3)(.33,.4)(0,.3)(0,0)}
\linethickness{0.7pt}
\qbezier(0,0.03)(3.45,1.3)(6.9,.03)\qbezier(0.2,0.03)(1.9,.6)(3.4,.03)
\put(0,-.03){$\underbrace{\qquad\qquad}_p$}
\put(2.15,-.03){$\underbrace{\qquad\qquad}_{j}$}
\put(3.55,-.03){$\underbrace{\quad\quad}_{s-j}$}
\put(5.7,-.03){$\underbrace{\quad\qquad}_{r}$}
\end{pspicture}
\end{center}
\end{figure}

Thus, this contribution gives $y{z^{\prime}}^{r+2}B_{r,s}(x,y)\sum_{j=1}^{s}{%
z^{\prime}}^{s-j}B_{r,s}(x,y;p-1,j)$. Putting together the two disjoint
cases above, we obtain \emph{(ii)}, as required.

\item If $q\geq2$ and $p=1$ then the degree of $v$, the second vertex (which
is black), is either zero or one. The first contribution gives $z^{\prime
}B_{r,s}(x,y;1,q-1)$. In the second case, there exists a white vertex
connected to $v$. Then $\pi$ can be represented as

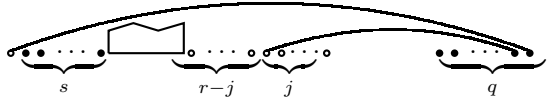
\begin{figure}[ht]
\begin{center}
\begin{pspicture}(0,-.3)(7,.8)
\pscircle(0,0){0.05}\multips(0.2,0)(0.2,0){2}{\pscircle*(0,0){0.05}}\put(.6,0){$\ldots$}\pscircle*(1.2,0){0.05}
\multips(2.4,0)(0.2,0){1}{\pscircle(0,0){0.05}}\put(2.6,0){$\ldots$}\pscircle(3.2,0){0.05}
\multips(3.4,0)(0.2,0){2}{\pscircle(0,0){0.05}}\put(3.7,0){$\ldots$}\pscircle(4.2,0){0.05}
\multips(5.7,0)(0.2,0){2}{\pscircle*(0,0){0.05}}\put(6.1,0){$\ldots$}\pscircle*(6.7,0){0.05}\pscircle*(6.9,0){0.05}
\multips(1.3,0)(0,0){1}{\psline(0,0)(1,0)(1,.4)(.66,.3)(.33,.4)(0,.3)(0,0)}
\linethickness{0.7pt}
\qbezier(0,0.03)(3.45,1.3)(6.9,.03)\qbezier(6.7,0.03)(5.1,.6)(3.4,.03)
\put(0.15,-.05){$\underbrace{\qquad\quad\,}_s$}
\put(2.15,-.05){$\underbrace{\qquad\quad\,\,}_{r-j}$}
\put(3.35,-.05){$\underbrace{\quad\quad}_{j}$}
\put(5.7,-.05){$\underbrace{\qquad\qquad}_{q}$}
\end{pspicture}
\end{center}
\end{figure}

Thus, this contribution gives $y{z^{\prime}}^{s+2}B_{r,s}(x,y)\sum_{j=1}^{r}{%
z^{\prime}}^{r-j}B_{r,s}(x,y;j,q-1)$. Putting together these two disjoint
cases, we get $(ii)$, as requested.

\item Let $p,q\geq2$. We consider the following two cases corresponding to
the possible degrees of $v$, the vertex (which is white) before the last
vertex. The contribution of the case in which the degree of $v$ is zero
gives $z^{\prime}B_{r,s}(x,y;p-1,q)$. If the degree of $v$ is one, then
there exists a black vertex $w$ connected to $v$. Then there are two
possibilities: $w$ is one of the first $q-1$ black vertices or is not one of
those. The contribution of the first case gives $y{z^{\prime}}%
^{2}\sum_{j=1}^{q-1}{z^{\prime}}^{q-1-j}B_{r,s}(x,y;p-1,j)$. The
contribution of the second case (similar to the case $p\geq2$ and $q=1$)
gives $y{z^{\prime}}^{2}B_{r,s}(x,y;q-1)\sum _{j=1}^{s}{z^{\prime}}%
^{s-j}B_{r,s}(x,y;p-1,j)$. Putting together the two disjoint cases above, we
obtain \emph{(iv)}, as claimed.
\end{itemize}
\end{proof}

Lemma~\ref{lemaa1} and Lemma~\ref{lemaa2} together with Eq. (\ref{eqa1})
give a (nonlinear) system of equations in the variables $B_{r,s}(x,y)$, $%
B_{r,s}(x,y;t)$ ($t=0,1,\ldots,s$), and $B_{r,s}(x,y;p,q)$ ($p=1,2,\ldots,r$
and $q=1,2,\ldots,s$). We solve this system for several interesting cases.

\begin{theorem}
Let $r\geq1$. Then
\begin{equation*}
B_{r,1}(x,y)=B_{1,r}(x,y)=(1+xB_{1,r}(x,y))(1+xyB_{1,r}(x,y))^{r}.
\end{equation*}
Moreover, for all $n\geq0$,
\begin{equation*}
\begin{array}{l}
\mathcal{NC}\biggl((a^{r}a^{\dag})^{n}\biggr)=\sum\limits_{j=0}^{n}\frac {1}{%
n+1}\binom{n+1}{j+1}\binom{rn+r}{j}(a^{\dag})^{n-j}a^{rn-j}, \\
\mathcal{NC}\biggl((a(a^{\dag})^{r})^{n}\biggr)=\sum\limits_{j=0}^{n}\frac {1%
}{n+1}\binom{n+1}{j+1}\binom{rn+r}{j}(a^{\dag})^{rn-j}a^{n-j}.%
\end{array}%
\end{equation*}
\end{theorem}

The above theorem can be proved combinatorially as described in Section~\ref%
{secc}. Another application of Lemma~\ref{lemaa1} and Lemma~\ref{lemaa2} is
the next observation.

\begin{theorem}
The generating function $B_{2,2}(x,y)$ satisfies
\begin{equation*}
\begin{array}{ll}
B_{2,2}(x,y) & =1+x(1+y)^{2}B_{2,2}(x,y)+2xy(1+x(1+y)+xy^{2})B_{2,2}^{2}(x,y)
\\[4pt]
& \qquad\qquad\qquad\qquad%
\qquad+x^{2}y^{2}(x(1+y)^{2}-1)B_{2,2}^{3}(x,y)+x^{4}y^{4}B_{2,2}^{4}(x,y).%
\end{array}%
\end{equation*}
\end{theorem}

\section{Noncrossing normal ordering form of $(a^{r}+(a^{\dagger})^{s})^{n}$}

Let us denote by $\mathcal{W}_{rs}(n)$ the set of all the linear
representations of the noncrossing contractions of $(a^{r}+(a^{%
\dagger})^{s})^{n}$. For each linear representation $\pi\in\mathcal{W}%
_{rs}(n)$, define $w(\pi)$ (resp. $e(\pi)$) to be the number of white
vertices (resp. edges) in $\pi$ and. Let $A_{r,s}(x,y,z)$ be the
generating function for the number of linear representations $\pi\in%
\mathcal{W}_{rs}(n)$ with exactly $m$ edges and $d$ white vertices, that is,
\begin{equation*}
A_{r,s}(x,y,z)=\sum_{n\geq0}\sum_{\pi\in\mathcal{W}_{rs}(n)}x^{n}y^{e(\pi
)}z^{w(\pi)}.
\end{equation*}
Hence, the noncrossing normally ordered form of $(a^{r}+(a^{\dag})^{s})^{n}$
is given by
\begin{equation*}
\mathcal{NC}\biggl((a^{r}+(a^{\dag})^{s})^{n}\biggr)=\sum_{i\geq0}\sum
_{j=0}^{i}(\mbox{the coefficient of }x^{n}y^{j}z^{i}\mbox{ in
}A_{r,s}(x,y,z))(a^{\dag})^{n-ri-j}a^{i-j}.
\end{equation*}
In order to find the noncrossing normally ordered form of $(a^{r}+(a^{\dag
})^{s})^{n}$, it is enough to find an explicit formula for the generating
function $A_{r,s}(x,y,z)$. In this section, we present a nonlinear system of
equations whose solution gives an explicit formula for $A_{r,s}(x,y,z)$. We
write
\begin{equation}
A_{r,s}(x,y,z)=1+xz^{r}A_{r,s}(x,y,z)+A_{r,s}(x,y,z;s),  \label{eqb1}
\end{equation}
where $A_{r,s}(x,y,z;t)$ is the generating function for all the linear
representations $\pi\in\mathcal{W}_{rs}(n)$ such that the canonical
sequential form of $\pi$ starts with $12\ldots t$. Applying a similar
argument as in the proof of Lemma~\ref{lemaa1}, we have a recurrence
relation for the sequence $A_{r,s}(x,y,z;t)$.

\begin{lemma}
\label{lembb1} Let $z^{\prime}=x^{\frac{1}{s}}$ and ${z^{\prime\prime}}=x^{%
\frac{1}{r}}$. For all $t=1,2,\ldots,s$,
\begin{equation*}
A_{r,s}(x,y,z;t)=z^{\prime}A_{r,s}(x,y,z;t-1)+A_{r,s}(x,y,z)%
\sum_{j=1}^{r}(z^{\prime\prime}z)^{r-j}A_{r,s}(x,y,z;j,t)
\end{equation*}
with the initial condition $A_{r,s}(x,y,z;0)=A_{r,s}(x,y,z)$, where $%
A_{r,s}(x,y;a,b)$ is the generating function for the number of linear
representations $\pi\in\mathcal{W}_{rs}(n)$, such that $\pi$ starts with $b$
black vertices, ends with $a$ white vertices and there is an edge between
the first black vertex and last white vertex.
\end{lemma}

\begin{proof}
Let $\pi$ be any linear representation in $\mathcal{W}_{rs}(n)$ such that $%
\pi$ starts with $t$ black vertices. Let us write an equation for $%
B_{r,s}(x,y;t)$. The first black vertex has degree either zero or one. The
contribution of the first case is $z^{\prime}A_{r,s}(x,y,z;t-1)$. Now let us
consider the second case, that is, the first black vertex has degree one.

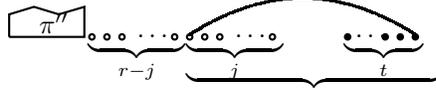
\begin{figure}[ht]
\begin{center}
\begin{pspicture}(0,-.2)(5.5,.6)
\psline(0,0)(1,0)(1,.4)(.66,.3)(.33,.4)(0,.3)(0,0)
\put(0.4,.05){$\pi''$}
\multips(0,0)(0.2,0){3}{\pscircle(1.1,0){0.05}}\put(1.7,0){$\ldots$}
\multips(1.1,0)(0.2,0){4}{\pscircle(1.1,0){0.05}}\put(3,0){$\ldots$}\pscircle(3.5,0){0.05}
\pscircle*(4.5,0){0.05}\put(4.6,0){$\ldots$}\multips(3.9,0)(0.2,0){3}{\pscircle*(1.1,0){0.05}}
\linethickness{0.7pt}\qbezier(5.4,0.03)(3.9,1)(2.4,.03)
\put(4.45,-.05){$\underbrace{\qquad\quad}_t$}
\put(1.05,-.05){$\underbrace{\qquad\quad\,\,\,\,}_{r-j}$}
\put(2.35,-.05){$\underbrace{\qquad\quad\,\,\,\,}_{j}$}
\put(2.35,-.45){$\underbrace{\qquad\quad\qquad\quad\qquad\quad\,\,\,\,}_{\pi'}$}
\end{pspicture}
\end{center}
\par
\label{figlemaa11}
\end{figure}

Thus our contraction $\pi $ can be written as $\pi =\pi ^{\prime
}(p+1)^{\prime }\ldots q^{\prime }\pi ^{\prime \prime }$, see the above
figure, where $\pi ^{\prime }$ starts with $t$ black vertices and ends with $%
j$ white vertices; there is an edge between the first back vertex of $\pi
^{\prime }$ and the last white vertex of $\pi ^{\prime }$; $\pi ^{\prime }$
is followed by $q-p$ white vertices such that $q-p+j=r$. Since we are
interested in the noncrossing contractions, there are no edges from $\pi
^{\prime }(p+1)^{\prime }\ldots q^{\prime }$ and $\pi ^{\prime \prime }$.
Therefore, the contribution of this second case gives
\begin{equation*}
A_{r,s}(x,y,z)\sum_{j=1}^{r}(zz^{\prime \prime })^{r-j}A_{r,s}(x,y,z;j,t).
\end{equation*}%
Adding the above disjoint cases we obtain the recurrence relation. From the
definitions, we obtain that $A_{r,s}(x,y,z;0)=A_{r,s}(x,y,z)$, as required.
\end{proof}

As we see in Lemma~\ref{lembb1}, to find a formula for the generating
function $A_{r,s}(x,y,z)$, we need to find a recurrence relation for the
generating functions $A_{r,s}(x,y,z;p,q)$. This can be done by using a
similar techniques as in the proof of Lemma~\ref{lemaa2}, which gives a
recurrence relation for the sequence $A_{r,s}(x,y,z;p,q)$.

\begin{lemma}
\label{lembb2} Let $z^{\prime}=x^{\frac{1}{s}}$ and ${z^{\prime\prime}}=x^{%
\frac{1}{r}}$, $p=2,3,\ldots,r$, and $q=2,3,\ldots,s$. Then the following
holds:

\textrm{(i)} $A_{r,s}(x,y,z;1,q)=yzz^{\prime}z^{\prime%
\prime}A_{r,s}(x,y,z;q-1)$.

\textrm{(ii)} $%
\begin{array}{ll}
A_{r,s}(x,y,z;p,q)= & yzz^{\prime}z^{\prime\prime}A_{r,s}(x,y,z;p-1,q)
+yz^{\prime}z^{\prime\prime}\sum\limits_{j=1}^{q-1}{z^{\prime}}%
^{q-1-j}A_{r,s}(x,y,z;p-1,j) \\
& \qquad\qquad\qquad
+yzz^{\prime}z^{\prime\prime}A_{r,s}(x,y,z;q-1)\sum\limits_{j=1}^{s}{%
z^{\prime}}^{s-j}A_{r,s}(x,y,z;p-1,j).%
\end{array}
$
\end{lemma}

\begin{proof}
Let $\pi$ be any linear representations in $\mathcal{W}_{rs}(n)$ such that
the last $p$ vertices of $\pi$ are white, the first $q$ vertices of $\pi$
are black, and there is an edge between the first black vertex and the last
white vertex. The generating function for the number of such linear
representations $\pi$, respect to the number vertices, the number of white
vertices and the number edges in $\pi$, is given by $A_{r,s}(x,y,z;p,q)$.
Now, let us write an equation for $A_{r,s}(x,y,z;p,q)$ for each of the
following two cases:

\begin{itemize}
\item If $p=1$ and $q\geq1$ then it is not hard to see from the definitions
that $A_{r,s}(x,y,z;1,q)=yzz^{\prime}z^{\prime\prime}A_{r,s}(x,y,z;q-1)$.

\item Let $p\geq2$. Let us consider the following two cases that correspond
to the degree of $v$, the vertex before the last vertex (which is white).
The contribution of the case when the degree of $v$ is zero gives $%
yzz^{\prime }z^{\prime\prime}A_{r,s}(x,y,z;p-1,q)$. If degree of $v$ is one
then there exists a black vertex $w$ connected by an edge with $v$. There are
two possibilities: $w$ is one of the first $q-1$ black vertices or it is
not. Applying similar arguments as in the proof of Lemma~\ref{lemaa2}, we
can see that the contribution of the first case gives
\begin{equation*}
yz^{\prime}z^{\prime\prime}\sum_{j=1}^{q-1}{z^{\prime}}%
^{q-1-j}A_{r,s}(x,y,z;p-1,j)
\end{equation*}
and the contribution of the second case gives
\begin{equation*}
yzz^{\prime}z^{\prime\prime}A_{r,s}(x,y,z;q-1)\sum_{j=1}^{s}{z^{\prime}}%
^{s-j}A_{r,s}(x,y,z;p-1,j).
\end{equation*}
Together, the two case above give \emph{(ii)}, as claimed.
\end{itemize}
\end{proof}

Hence, Lemma~\ref{lembb1} and Lemma~\ref{lembb2} together with (\ref{eqb1})
gives a (nonlinear) system of equations in the variables $A_{r,s}(x,y,z)$, $%
A_{r,s}(x,y,z;t)$ ($t=0,1,\ldots,s$), and $A_{r,s}(x,y,z;p,q)$ ($%
p=1,2,\ldots ,r$ and $q=1,2,\ldots,s$).

\begin{theorem}
The generating function $A=A_{r,1}(x,y,z)$ satisfies
\begin{equation*}
A=1+x(1+z^{r})A+x^{2}yz^{r}A^{r}\frac{1-x^{r}y^{r}(1+A)^{r})}{1-xy(1+A)}.
\end{equation*}
\end{theorem}

For example, when $r=1$, the above theorem gives
\begin{equation*}
A_{1,1}(x,y,z)=\frac{1-xz-x-\sqrt{(1-x-xz)^{2}-4x^{2}yz}}{2x^{2}yz}.
\end{equation*}
Using the fact that $\frac{1-\sqrt{1-4x}}{2x}=\sum_{n\geq0}c_{n}x^{n}$,
where $c_{n}=\frac{1}{n+1}\binom{2n}{n}$ the $n$-th Catalan number, we have
\begin{equation*}
A_{1,1}(x,y,z)=\sum_{n\geq0}\sum_{j=0}^{n}\sum_{i=j}^{n-j}c_{j}\binom{n}{i+j}%
\binom{i+j}{2j}x^{n}y^{j}z^{i}.
\end{equation*}
Thus, the noncrossing normally ordered form of $(a+a^{\dag})^{n}$ is given
by
\begin{equation}  \label{eqsst}
\mathcal{NC}\biggl((a+a^{\dag})^{n}\biggr)=\sum_{j=0}^{n}%
\sum_{i=j}^{n-j}c_{j}\binom{n}{i+j}\binom{i+j}{2j}(a^{\dag})^{n-i-j}a^{i-j}.
\end{equation}

Another application of Lemma~\ref{lembb1} and Lemma~\ref{lembb2} with $r=1$
and $s\geq1$ is that the generating function $A_{1,s}(x,y,z)$ satisfies the
equation
\begin{equation}  \label{eqrrf}
A_{1,s}(x,y,z)=1+xzA_{1,s}(x,y,z)+xA_{1,s}(x,y,z)(1+xyzA_{1,s}(x,y,z))^{s}.
\end{equation}
By Lagrange inversion formula \cite{wilf} on the above equation, we obtain
that
\begin{equation*}
A_{1,s}(x,y,z)=\sum_{n\geq0}\sum_{j=0}^{n}\sum_{i=j}^{n}\frac{1}{n+1}\binom{%
n+1}{j+1}\binom{n-j}{n-i}\binom{s(n-i)}{j}x^{n}y^{j}z^{i},
\end{equation*}
which leads to the following result.

\begin{theorem}
\label{thaa8} The normally ordered form of $(a+(a^{\dag})^{s})^{n}$ is given
by
\begin{equation*}
\mathcal{NC}\biggl((a+(a^{\dag})^{s})^{n}\biggr)=\sum_{j=0}^{n}\sum_{i=j}^{n}%
\frac{1}{n+1}\binom{n+1}{j+1}\binom{n-j}{n-i}\binom{s(n-i)}{j}(a^{\dag
})^{s(n-i)-j}a^{i-j}.
\end{equation*}
\end{theorem}

We remark that Theorem~\ref{thaa8} for $s=1$ gives
\begin{equation*}
\begin{array}{l}
\mathcal{NC}\biggl((a+a^{\dag})^{n}\biggr) \\
\qquad=\sum\limits_{j=0}^{n}\sum\limits_{i=j}^{n-j}\frac{1}{n+1}\binom{n+1}{%
j+1}\binom{n-j}{n-i}\binom{n-i}{j}(a^{\dag})^{n-i-j}a^{i-j}=\sum%
\limits_{j=0}^{n}\sum\limits_{i=j}^{n-j}\frac{n!}{j!(j+1)!(i-j)!(n-i-j)!}%
(a^{\dag})^{n-i-j}a^{i-j} \\
\qquad=\sum\limits_{j=0}^{n}\sum\limits_{i=j}^{n-j}\frac{(2j)!}{j!(j+1)!}%
\frac{n!}{(i+j)!(n-i-j)!}\frac{(i+j)!}{(2j)!(i-j)!}(a^{%
\dag})^{n-i-j}a^{i-j}=\sum\limits_{j=0}^{n}\sum\limits_{i=j}^{n-j}c_{j}%
\binom{n}{i+j}\binom{i+j}{2j}(a^{\dag})^{n-i-j}a^{i-j},%
\end{array}%
\end{equation*}
as described in \eqref{eqsst}.

\section{Two particular cases, $k$-ary trees, and lattice paths}

\label{secc} In this section, we relate our linear representations $\mathcal{%
V}_{rs}(n)$ and $\mathcal{W}_{rs}(n)$, for several particular cases of $r$
and $s$, to different combinatorial structure, such as $k$-ary trees,
lattice paths, and Dyck paths (these notions will be defined below).

\subsection{The case $(a^{r}a^{\dagger})^{n}$ and $k$-ary trees}

Our goal is to show that the number of noncrossing contractions of $%
(a^{r}a^{\dag})^{n}$ is counted by the generalized Catalan numbers $C_{n,k}$%
, defined by $C_{n,k}={\frac{1}{{(k-1)n+1}}}{\binom{kn}{n}}$. In our
approach, we give a recursive construction of the set $\mathcal{V}_{r1}(n)$.
Intuitively, for a noncrossing contraction $\pi=\pi_{1}\pi_{2}\cdots\pi_{n}$
in $\mathcal{V}_{r1}(n)$, we may obtain a contraction in $\mathcal{V}%
_{r1}(n-1)$. Then we need to keep track of all possible ways to recover a
noncrossing contraction in $\mathcal{V}_{r1}(n)$ from a smaller noncrossing
contraction. In the recursive generation of noncrossing contractions with $n$
vertices, one is often concerned with the edges whose points have the label $%
1$ in the canonical sequential form. However, for the purpose of this paper,
we consider edges whose first point is labeled by $1$. We denote by $E_{\pi}$
the edge $(1,j)$. In general, we use the notation $E_{i}$ to denote the edge
with first point $i$. In this sense, $E_{\pi}=E_{1}$. Also, we make use of
the ordered set $F_{\pi}$ of all white vertices $j-1,j-2,\ldots,i+1$, where $%
i$ is maximal, $i<j$ and $i$ is a black vertex.

An edge $(i,j)$ is said to \emph{cover} the edge $(i^{\prime},j^{\prime})$
if and only if $i<i^{\prime}<j^{\prime}<j$. We have the following lemma on
the structure of the noncrossing contractions of $(a^{r}a^{\dag})^{n}$. The
lemma is straightforward to prove.

\begin{lemma}
\label{lem1} Let $\pi$ be any noncrossing contraction in the set $\mathcal{V}%
_{r1}(n)$. Then the canonical subsequential form of $\pi$ is
\begin{equation*}
\begin{array}{l}
a_{1}(a_{1}+1)^{\prime}\ldots(a_{1}+r)^{\prime}\pi^{(1)}\cdots
a_{m-1}(a_{m-1}+1)^{\prime}\ldots(a_{m-1}+r)^{\prime}\pi^{(m-1)} \\
\qquad\quad
a_{m}\pi^{(m)}(a_{m}+s_{0})^{\prime}\ldots(a_{m}+s_{1}-1)^{%
\prime}a_{m}(a_{m}+s_{1})^{\prime}\ldots(a_{m}+s_{2}-1)^{\prime} \\
\qquad\qquad\qquad\quad
a_{m-1}\cdots(a_{m}+s_{m-1})^{\prime}\ldots(a_{m}+s_{m}-1)^{%
\prime}a_{1}(a_{m}+s_{m})^{\prime}\ldots(a_{m}+s_{m+1}-1)^{\prime}%
\pi^{(m+1)},%
\end{array}%
\end{equation*}
where $a_{1}=1$, $a_{i+1}-a_{i}>r$, $s_{i+1}\geq s_{i}\geq0$, $\pi^{(i)}\in%
\mathcal{V}_{r1}(a_{i+1}-a_{i}-r-1)$, $i=1,2,\ldots,m-1$, $\pi^{(m)}$ is
either empty ($s_{0}=1$) or $\pi^{(m)}=(a_{m}+1)^{\prime}\ldots(a_{m}+r)^{%
\prime}\theta(a_{m}+s_{0}-1)$ with $\theta\in\mathcal{V}_{r1}(s_{0}-2-r)$,
and $\pi^{(m+1)}\in\mathcal{V}_{r1}(n-a_{m+1}-s_{0}+1)$. In other words,
there exist $m$ edges, say $E_{i_{1}}=E_{\pi},E_{i_{2}},\ldots ,E_{i_{m}}$,
such that the linear representation of $\pi$ is either
\begin{figure}[ht]
\begin{center}
\begin{pspicture}(-1,0)(14,1)
\multips(0,0)(.25,0){3}{\pscircle(0,0){0.05}}\put(.7,0){$...$}
\multips(0,0)(.25,0){3}{\pscircle(1.2,0){0.05}}\put(1.9,0){$...$}
\multips(0,0)(.25,0){3}{\pscircle(2.4,0){0.05}}\put(3.1,0){$...$}
\multips(0,0)(.25,0){3}{\pscircle(3.6,0){0.05}}
\multips(0,0)(2.5,0){3}{\pscircle(8,0){0.05}\pscircle(8.25,0){0.05}\pscircle(8.75,0){0.05}\pscircle*(9,0){0.05}}
\pscircle*(6.5,0){0.05}
\put(8.35,0){$...$}\put(10.85,0){$...$}\put(13.35,0){$...$}\put(6.7,0.1){$...$}
\linethickness{0.7pt}
\qbezier(.25,.03)(7.1,2.3)(14,.03)\put(12.7,.47){\footnotesize$E_{i_1}$}
\qbezier(1.45,.03)(5.8,2)(11.5,.03)\put(10.2,.47){\footnotesize$E_{i_2}$}
\qbezier(2.65,.03)(4.5,1.7)(9,.03)\put(7.8,.47){\footnotesize$E_{i_3}$}
\qbezier(3.85,.03)(4.9,1.4)(6.5,.03)\put(5.9,.47){\footnotesize$E_{i_{m}}$}
\put(-.55,-.05){$\underbrace{\qquad\qquad\qquad\qquad\qquad\qquad\qquad\qquad}_r$}
\put(8,-.05){$\underbrace{\qquad\,\,\,}_r$}
\put(10.5,-.05){$\underbrace{\qquad\,\,\,}_r$}
\put(13,-.05){$\underbrace{\qquad\,\,\,}_r$}
\multips(0,0)(-2.5,0){3}{\psline[linewidth=.01](12,0)(12.6,0)(12.6,.2)(12,.2)(12,0)}
\put(4.2,0){$...$}\pscircle(4.5,0){0.05}
\put(-.36,0){$...$}\pscircle(-.5,0){0.05}
\multips(-13.3,0)(0,0){1}{\psline[linewidth=.01](11.8,0)(12.6,0)(12.6,.2)(11.8,.2)(11.8,0)}
\end{pspicture}
\end{center}
\caption{First factorization}
\label{figfact1}
\end{figure}
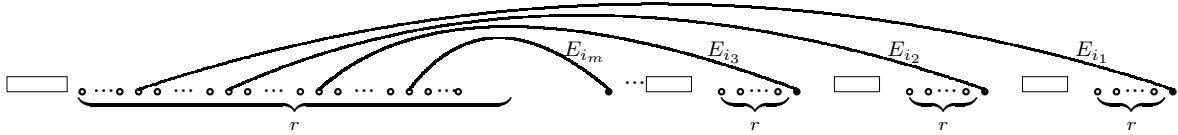
\noindent or
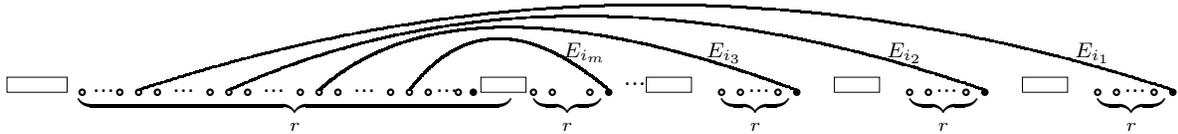
\begin{figure}[ht]
\begin{center}
\begin{pspicture}(-1,0)(14,1)
\multips(0,0)(.25,0){3}{\pscircle(0,0){0.05}}\put(.7,0){$...$}
\multips(0,0)(.25,0){3}{\pscircle(1.2,0){0.05}}\put(1.9,0){$...$}
\multips(0,0)(.25,0){3}{\pscircle(2.4,0){0.05}}\put(3.1,0){$...$}
\multips(0,0)(.25,0){3}{\pscircle(3.6,0){0.05}}
\multips(-2.5,0)(2.5,0){4}{\pscircle(8,0){0.05}\pscircle(8.25,0){0.05}\pscircle(8.75,0){0.05}\pscircle*(9,0){0.05}}
\pscircle*(4.7,0){0.05}
\put(8.35,0){$...$}\put(10.85,0){$...$}\put(13.35,0){$...$}\put(6.7,0.1){$...$}
\linethickness{0.7pt}
\qbezier(.25,.03)(7.1,2.3)(14,.03)\put(12.7,.47){\footnotesize$E_{i_1}$}
\qbezier(1.45,.03)(5.8,2)(11.5,.03)\put(10.2,.47){\footnotesize$E_{i_2}$}
\qbezier(2.65,.03)(4.5,1.7)(9,.03)\put(7.8,.47){\footnotesize$E_{i_3}$}
\qbezier(3.85,.03)(4.9,1.4)(6.5,.03)\put(5.9,.47){\footnotesize$E_{i_m}$}
\put(-.55,-.05){$\underbrace{\qquad\qquad\qquad\qquad\qquad\qquad\qquad\qquad}_r$}
\put(5.5,-.05){$\underbrace{\qquad\,\,\,}_r$}
\put(8,-.05){$\underbrace{\qquad\,\,\,}_r$}
\put(10.5,-.05){$\underbrace{\qquad\,\,\,}_r$}
\put(13,-.05){$\underbrace{\qquad\,\,\,}_r$}
\multips(0,0)(-2.5,0){3}{\psline[linewidth=.01](12,0)(12.6,0)(12.6,.2)(12,.2)(12,0)}
\multips(-7,0)(0,0){1}{\psline[linewidth=.01](11.8,0)(12.4,0)(12.4,.2)(11.8,.2)(11.8,0)}
\put(4.2,0){$...$}\pscircle(4.5,0){0.05}
\put(-.36,0){$...$}\pscircle(-.5,0){0.05}
\multips(-13.3,0)(0,0){1}{\psline[linewidth=.01](11.8,0)(12.6,0)(12.6,.2)(11.8,.2)(11.8,0)}
\end{pspicture}
\end{center}
\caption{Second factorization}
\label{figfact2}
\end{figure}

where each edge $E_{i_{j}}$ covers the edge $E_{i_{j+1}}$, such that the end
points of $E_{i_{1}},...,E_{i_{m}}$ are white vertices, $v_{1},...,v_{m}$,
and there is no black vertex between $v_{i}$ and $v_{j}$, for any $i$ and $j$%
.
\end{lemma}

A \emph{$k$-ary tree} is a directed tree in which each vertex has degree $0$
or $k$ (see, \emph{e.g.}, \cite{St}). The number of $k$-ary trees with $n$
vertices is counted by the \emph{$k$-ary numbers}, defined by $\frac{1}{kn+1}%
\binom{kn+1}{n}$, for any positive integers $k$ and $n$. Let $\mathcal{T}%
_{r,n}$ be the set of $r$-ary tree with $n$ nodes. We denote by $%
T^{1},\ldots,T^{r}$ the children of its root (from right to left). Now we are ready to define a bijection $\Phi$ recursively. Firstly, the empty
contraction maps to the empty $(r+1)$-ary tree, which gives the bijection $%
\Phi \colon\mathcal{V}_{r1}(0)\mapsto\mathcal{T}_{r+1,0}$. Define $%
F_{\pi}^{\prime }=\{k_{1},k_{2},\ldots,k_{m}\}$ to be an ordered subset
(that is, $k_{1}>k_{2}>\ldots>k_{m}$) of $F_{\pi}$, such that the node $%
k_{i} $ is the end point of the edge $E_{i_{m+1-i}}$. Define the minimal vertex
of $F_{\pi}$ by $k_{0}$, that is, $k_{0}=\min_{i\in F_{\pi}}i$. Suppose we
have defined the bijection $\Phi\colon\mathcal{V}_{r1}(m)\mapsto\mathcal{T}%
_{r+1,m}$ for all $m<n$. For $\pi\in\mathcal{V}_{r1}(n)$, according to the
factorizations of the contraction $\pi$ as described in Lemma~\ref{lem1},
there are two cases, for all $m\geq1$:

\begin{itemize}
\item If $\pi^{(m)}=\emptyset$ (see Figure~\ref{figfact1}) then we define
the $(k_{m}-k_{0}+1)$-th child of $T$ to be $T^{k_{1}-k_{0}+1}=\Phi(%
\pi^{(m+1)})$, and the $(k_{i}-k_{0}+1)$-th child of $T$ to be $%
T^{k_{i}-k_{0}+1}=\Phi^{(}\pi^{(i-1)})$, for each $i=2,3,\ldots,m$.

\item If $\pi^{(m)}\neq\emptyset$ (see Figure~\ref{figfact2}) then define
the $(r+1)$-th child of $T$ to be $T^{r+1}=\Phi(\pi^{(m+1)})$, and the $%
(k_{i}-k_{0}+1)$-st child of $T$ to be $T^{k_{i}-k_{0}+1}=\Phi^{(}\pi^{(i)})$%
, for each $i=1,2,3,\ldots,m$.
\end{itemize}

In the case of $m=0$, we define our tree $T$ to be a $(r+1)$-ary tree with
root having only one child, which is $T^{r+1}=\Phi(\pi^{(1)})$. We can see
that if there is no child $T^{r+1}$ of the root of the $(r+1)$-ary tree $T$,
then this tree corresponds to a noncrossing contraction with a factorization
as described in Figure~\ref{figfact1}; otherwise, the tree corresponds to a
noncrossing contraction with a factorization as described in Figure~\ref%
{figfact2}. By induction on the length of the noncrossing contractions and
the unique construction of the map $\Phi$, we have that $\Phi$ is
invertible. It follows that is a bijection between the set of the
noncrossing contractions $\mathcal{V}_{r1}(n)$ and the set $\mathcal{T}%
_{r+1,n}$ of $(r+1)$-ary trees. Thus, we have shown the following result.

\begin{theorem}
There is a bijection between the set of noncrossing contractions in $%
\mathcal{V}_{r1}(n)$ and the set $\mathcal{T}_{r+1,n}$ of $(r+1)$-ary trees.
\end{theorem}

Let $G_{r}(x,y)$ be the generating function for the number of noncrossing
contractions in $\mathcal{V}_{r1}(n)$ with exactly $m$ arcs, that is,
\begin{equation*}
G_{r}(x,y)=\sum_{n\geq0}\sum_{\pi\in\mathcal{V}_{r1}(n)}x^{n}y^{\#arcs\ in\
\pi}.
\end{equation*}
Then the bijection $\phi$ gives that $%
G_{r}(x,y)=(1+xG_{r}(x,y))(1+xyG_{r}(x,y))^{r}$. By Lagrange inversion
formula \cite{wilf} on the above equation, we obtain that
\begin{equation*}
G_{r}(x,y)=\sum_{n\geq1}\frac{x^{n-1}}{n}\sum_{i=0}^{n-1}\binom{n}{i}\binom{%
rn}{n-1-i}y^{n-1-i}.
\end{equation*}
As a consequence, we have the following theorem.

\begin{theorem}
The noncrossing normally ordered from of $(a^{r}a^{\dag})^{n}$ is given by
\begin{equation*}
\mathcal{NC}\biggl((a^{r}a^{\dag})^{n}\biggr)=\sum_{j=0}^{n}\frac{1}{n+1}%
\binom{n+1}{j+1}\binom{rn+r}{j}(a^{\dag})^{n-j}a^{rn-j}.
\end{equation*}
\end{theorem}

Using similar arguments as in the construction of the bijection $\Phi$, one
can obtain a bijection between the set of the noncrossing contractions of $%
(a(a^{\dag})^{r})^{n}$ and the set of $(r+1)$-ary trees with $n$ nodes.

\begin{theorem}
\label{th33} 
The noncrossing normally ordered from of $(a(a^{\dag})^{r})^{n}$ is given by
\begin{equation*}
\mathcal{NC}\biggl((a(a^{\dag})^{r})^{n}\biggr)=\sum_{j=0}^{n}\frac{1}{n+1}%
\binom{n+1}{j+1}\binom{rn+r}{j}(a^{\dag})^{rn-j}a^{n-j}.
\end{equation*}
\end{theorem}

\subsection{The case $(a+(a^{\dagger})^{2})^{n}$ and lattice paths}

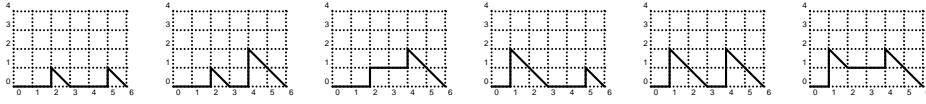
\begin{figure}[ht]
\begin{center}
\begin{pspicture}(0,0)(2,1)
\psgrid[unit=0.25,subgriddiv=1,griddots=5,gridlabels=3pt](0,0)(6,4)
\psline(0,0)(.5,0)(.5,.25)(.75,0)(1.25,0)(1.25,.25)(1.5,0)
\end{pspicture}
\begin{pspicture}(0,0)(2,1)
\psgrid[unit=0.25,subgriddiv=1,griddots=5,gridlabels=3pt](0,0)(6,4)
\psline(0,0)(.5,0)(.5,.25)(.75,0)(1,0)(1,.5)(1.5,0)
\end{pspicture}
\begin{pspicture}(0,0)(2,1)
\psgrid[unit=0.25,subgriddiv=1,griddots=5,gridlabels=3pt](0,0)(6,4)
\psline(0,0)(.5,0)(.5,.25)(1,.25)(1,.5)(1.5,0)
\end{pspicture}
\begin{pspicture}(0,0)(2,1)
\psgrid[unit=0.25,subgriddiv=1,griddots=5,gridlabels=3pt](0,0)(6,4)
\psline(0,0)(.25,0)(.25,.5)(.75,0)(1.25,0)(1.25,.25)(1.5,0)
\end{pspicture}
\begin{pspicture}(0,0)(2,1)
\psgrid[unit=0.25,subgriddiv=1,griddots=5,gridlabels=3pt](0,0)(6,4)
\psline(0,0)(.25,0)(.25,.5)(.75,0)(1,0)(1,.5)(1.5,0)
\end{pspicture}
\begin{pspicture}(0,0)(2,1)
\psgrid[unit=0.25,subgriddiv=1,griddots=5,gridlabels=3pt](0,0)(6,4)
\psline(0,0)(.25,0)(.25,.5)(.5,.25)(1,.25)(1,.5)(1.5,0)
\end{pspicture}\vspace{-10pt}
\end{center}
\caption{The set $\mathcal{P}_{2}$ of lattices paths}
\label{fig3}
\end{figure}

In this section, we present a bijection $\Phi$ between the set of
contractions $\mathcal{W}_{12}(n)$ of the word monomials in the expression $%
F(a,a^{\dagger })=(a+(a^{\dagger})^{2})^{n}$ and a special set $\mathcal{P}%
_{n}$ of $L$-lattice paths. These are lattice paths on $\mathbb{Z}^{2}$,
which are of length $4n$ and go from $(0,0)$ to $(3n,0)$. Moreover, the
paths never go below the $x$-axis with the steps $H=(2,1)$, $D=(1,-1)$ and $%
L=(1,2)$, and with no three consecutive $D$ steps (that is, there is no
triple $DDD$). The paths of length six are described in Figure~\ref{fig3}.

Let $P$ be a path in $\mathcal{P}_{n}$. Then, using a first return
decomposition (first return to $x$-axis) we obtain the factorization of $P$
as either
\begin{equation*}
HDP^{\prime},\,LDDP^{\prime},\, LDP^{\prime}HDDP^{\prime\prime},\,
HP^{\prime}HDDP^{\prime\prime}D,\,\mbox{or } LDP^{\prime}LDP^{\prime\prime
}HDDHDDP^{\prime\prime\prime},
\end{equation*}
where $P^{\prime},P^{\prime\prime},P^{\prime\prime\prime}$ are paths of
smaller length (see Figure~\ref{Fphi1}). On the basis of this observation,
the bijection $\Phi$ can be defined recursively. Firstly, the empty
contraction maps to the empty path, which gives the bijection $\Phi\colon%
\mathcal{W}_{12}(0)\mapsto\mathcal{P}_{0}$. Suppose we have defined the
bijection $\Phi\colon\mathcal{W}_{12}(m)\mapsto\mathcal{P}_{m}$ for all $m<n$%
. For $\pi\in\mathcal{W}_{12}(n)$, according to the factorizations of the
contraction $\pi$, there are five cases:

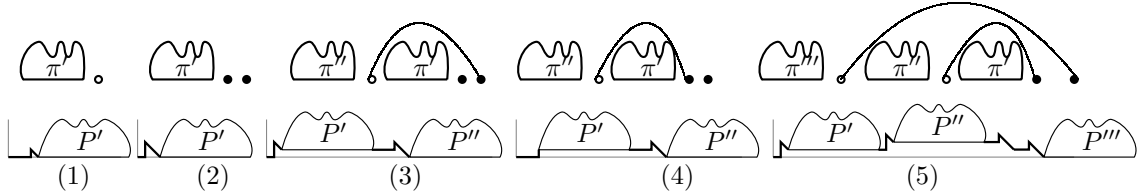
\begin{figure}[ht]
\begin{center}
\begin{pspicture}(-.1,0)(1.3,.7)
\setlength{\unitlength}{3mm} \linethickness{0.3pt}
\pscircle(1,0){.2}\pscurve(.8,0)(.7,.5)(.6,.3)(.5,.5)(.4,.3)(.2,.5)(0,0)(.01,0)(.8,0)
\put(1,0.15){$\pi'$}
\end{pspicture}
\begin{pspicture}(-.3,0)(1.3,.7)
\setlength{\unitlength}{3mm} \linethickness{0.3pt}
\pscircle*(1,0){.2}\pscircle*(1.25,0){.2}\pscurve(.8,0)(.7,.5)(.6,.3)(.5,.5)(.4,.3)(.2,.5)(0,0)(.01,0)(.8,0)
\put(1,.15){$\pi'$}
\end{pspicture}
\begin{pspicture}(-.7,0)(2.2,.7)
\setlength{\unitlength}{3mm} \linethickness{0.3pt}
\pscircle(0.8,0){.2}\pscircle*(2,0){.2}\pscircle*(2.25,0){.2}
\pscurve(1.8,0)(1.7,.5)(1.6,.3)(1.5,.5)(1.4,.3)(1.2,.5)(1,0)(1.01,0)(1.8,0)
\put(4.5,.15){$\pi'$}
\put(-.8,0){\pscurve(.8,0)(.7,.5)(.6,.3)(.5,.5)(.4,.3)(.2,.5)(0,0)(.01,0)(.8,0)}
\put(.3,.15){$\pi''$} \qbezier(7.5,0)(4.5,5)(2.5,.0)
\end{pspicture}
\begin{pspicture}(-.7,0)(2.1,.7)
\setlength{\unitlength}{3mm} \linethickness{0.3pt}
\pscircle(0.8,0){.2}\pscircle*(2,0){.2}\pscircle*(2.25,0){.2}
\pscurve(1.8,0)(1.7,.5)(1.6,.3)(1.5,.5)(1.4,.3)(1.2,.5)(1,0)(1.01,0)(1.8,0)
\put(4.5,.15){$\pi'$}
\put(-.8,0){\pscurve(.8,0)(.7,.5)(.6,.3)(.5,.5)(.4,.3)(.2,.5)(0,0)(.01,0)(.8,0)}
\put(.3,.15){$\pi''$} \qbezier(6.5,0)(5,5)(2.5,.0)
\end{pspicture}
\begin{pspicture}(-2.4,0)(3.3,.7)
\setlength{\unitlength}{3mm} \linethickness{0.3pt}
\pscircle(0.8,0){.2}\pscircle(-0.6,0){.2}\pscircle*(2,0){.2}\pscircle*(2.5,0){.2}
\pscurve(1.8,0)(1.7,.5)(1.6,.3)(1.5,.5)(1.4,.3)(1.2,.5)(1,0)(1.01,0)(1.8,0)
\put(4.5,.15){$\pi'$}
\put(-.8,0){\pscurve(.8,0)(.7,.5)(.6,.3)(.5,.5)(.4,.3)(.2,.5)(0,0)(.01,0)(.8,0)}
\put(-5.5,0){\pscurve(.8,0)(.7,.5)(.6,.3)(.5,.5)(.4,.3)(.2,.5)(0,0)(.01,0)(.8,0)}
\put(.3,.15){$\pi''$}\put(-4.5,.15){$\pi'''$}
\qbezier(6.6,0)(5.2,5)(2.5,.0)\qbezier(8.5,0)(3.2,6.8)(-2,.0)
\end{pspicture}
\par
\begin{pspicture}(0,0)(1.6,1)
\psline[linecolor=gray,linewidth=.01](0,0)(1.5,0)
\psline[linecolor=gray,linewidth=.01](0,0)(0,.5)
\psline[linecolor=black,linewidth=.03](0,0)(0.3,0)(0.3,.1)(0.4,.0)
\pscurve[linecolor=black,linewidth=.015](.4,0)(.7,.5)(.9,.4)(1,.5)(1.1,.4)(1.3,.5)(1.6,0)(1.59,0)(0.4,0)
\put(.9,.1){$P'$}
\end{pspicture}
\begin{pspicture}(0,0)(1.6,1)
\psline[linecolor=gray,linewidth=.01](0,0)(1.5,0)
\psline[linecolor=gray,linewidth=.01](0,0)(0,.5)
\psline[linecolor=black,linewidth=.03](0,0)(0.1,0)(0.1,.2)(0.3,.0)
\put(-.1,0){\pscurve[linecolor=black,linewidth=.015](.4,0)(.7,.5)(.9,.4)(1,.5)(1.1,.4)(1.3,.5)(1.6,0)(1.59,0)(0.4,0)
\put(.9,.1){$P'$}}
\end{pspicture}
\begin{pspicture}(0,0)(2.6,1)
\psline[linecolor=gray,linewidth=.01](0,0)(3,0)
\psline[linecolor=gray,linewidth=.01](0,0)(0,.5)
\psline[linecolor=black,linewidth=.03](0,0)(0.1,0)(0.1,.2)(.2,.1)
\put(-.2,.1){\pscurve[linecolor=black,linewidth=.015](.4,0)(.7,.5)(.9,.4)(1,.5)(1.1,.4)(1.3,.5)(1.6,0)(1.59,0)(0.4,0)
\put(.9,.1){$P'$}}
\put(1.4,.1){\psline[linecolor=black,linewidth=.03](0,0)(0.3,0)(0.3,.1)(.4,0)(.5,-.1)}
\put(1.5,0){\pscurve[linecolor=black,linewidth=.015](.4,0)(.7,.5)(.9,.4)(1,.5)(1.1,.4)(1.3,.5)(1.6,0)(1.59,0)(0.4,0)
\put(.9,.1){$P''$}}
\end{pspicture}
\begin{pspicture}(-.6,0)(2.4,1)
\psline[linecolor=gray,linewidth=.01](0,0)(3,0)
\psline[linecolor=gray,linewidth=.01](0,0)(0,.5)
\psline[linecolor=black,linewidth=.03](0,0)(0.3,0)(0.3,.1)
\put(-.1,.1){\pscurve[linecolor=black,linewidth=.015](.4,0)(.7,.5)(.9,.4)(1,.5)(1.1,.4)(1.3,.5)(1.6,0)(1.59,0)(0.4,0)
\put(.9,.1){$P'$}}
\put(1.5,.1){\psline[linecolor=black,linewidth=.03](0,0)(0.3,0)(0.3,.1)(.4,0)(.5,-.1)}
\put(1.6,0){\pscurve[linecolor=black,linewidth=.015](.4,0)(.7,.5)(.9,.4)(1,.5)(1.1,.4)(1.3,.5)(1.6,0)(1.59,0)(0.4,0)
\put(.9,.1){$P''$}}
\end{pspicture}
\begin{pspicture}(-.9,0)(4.9,1)
\psline[linecolor=gray,linewidth=.01](0,0)(4,0)
\psline[linecolor=gray,linewidth=.01](0,0)(0,.5)
\psline[linecolor=black,linewidth=.03](0,0)(0.1,0)(0.1,.2)(.2,.1)
\put(-.2,.1){\pscurve[linecolor=black,linewidth=.015](.4,0)(.7,.5)(.9,.4)(1,.5)(1.1,.4)(1.3,.5)(1.6,0)(1.59,0)(0.4,0)
\put(.9,.1){$P'$}}
\put(1.4,.1){\psline[linecolor=black,linewidth=.03](0,0)(0.1,0)(0.1,.2)(.2,.1)}
\put(1.2,.2){\pscurve[linecolor=black,linewidth=.015](.4,0)(.7,.5)(.9,.4)(1,.5)(1.1,.4)(1.3,.5)(1.6,0)(1.59,0)(0.4,0)
\put(.9,.1){$P''$}}
\psline[linecolor=black,linewidth=.03](2.8,.2)(3,.2)(3,.3)(3.2,.1)
\psline[linecolor=black,linewidth=.03](3.2,.1)(3.4,.1)(3.4,.2)(3.6,0)
\put(3.2,.0){\pscurve[linecolor=black,linewidth=.015](.4,0)(.7,.5)(.9,.4)(1,.5)(1.1,.4)(1.3,.5)(1.6,0)(1.59,0)(0.4,0)
\put(.9,.1){$P'''$}}
\end{pspicture}\vspace{-18pt}
\par
\begin{equation*}
\qquad\,(1)\qquad\qquad(2)\qquad\qquad\quad\quad(3)\qquad\qquad\qquad
\qquad\quad(4)\qquad\quad\quad\qquad\quad\quad(5)\qquad\qquad\qquad\qquad
\end{equation*}
\vspace{-30pt}
\end{center}
\caption{The bijection $\Phi$.}
\label{Fphi1}
\end{figure}

\begin{itemize}
\item[\emph{(i)}] The contraction $\pi$ starts with a white vertex, namely $%
\pi=1^{\prime}\pi^{\prime}\in\mathcal{W}_{12}(n)$. We define $\Phi(\pi)$ to
be the joint of the steps $HD$ and the path $P^{\prime}=\Phi(\beta)$, where $%
\beta_{i}=\pi_{i}^{\prime}-1$ (define $i^{\prime}-d=(i-d)^{\prime}$), for
each $i=1,2,\ldots,n-1$ (see Figure~\ref{Fphi1}(1)).

\item[\emph{(ii)}] The contraction $\pi$ starts with a black vertex, namely $%
\pi=1\pi^{\prime}\in\mathcal{W}_{12}(n)$. We define $\Phi(\pi)$ to be the
joint of the steps $LDD$ and the path $P^{\prime}=\Phi(\beta)$, where $%
\beta_{i}=\pi_{i}^{\prime}-1$, for each $i=1,2,\ldots,n-1$ (see Figure~\ref%
{Fphi1}(2)).

\item[\emph{(iii)}] The contraction $\pi$ starts with an arc followed by a
black vertex with degree zero, that is, $\pi=12\pi^{\prime}1\pi^{\prime%
\prime }\in\mathcal{W}_{12}(n)$, where in $\pi^{\prime}$ does not occur the
letter $2$. We define $\Phi(\pi)$ to be the joint of the step $LD$, the path
$P^{\prime}=\Phi(\beta^{\prime})$, the steps $HDD$, and the path $%
P^{\prime\prime}=\Phi(\beta^{\prime\prime})$, where, for each $i$, $\beta
_{i}^{\prime}=\pi_{i}^{\prime}-2$ and $\beta_{i}^{\prime\prime}=\pi
_{i}^{\prime\prime}-\max(2,\ell)$ such that $\ell$ is the maximal letter of $%
\pi^{\prime}$ (see Figure~\ref{Fphi1}(3)).

\item[\emph{(iv)}] The contraction $\pi$ starts with a black vertex of
degree zero followed by an arc, that is, $\pi=12\pi^{\prime}2\pi^{\prime%
\prime}\in\mathcal{W}_{12}(n)$, where in $\pi^{\prime\prime}$ does not occur
the letter $1$. We define $\Phi(\pi)$ to be the joint of the step $H$, the
path $P^{\prime}=\Phi(\beta^{\prime})$, the steps $HDD$, and the path $%
P^{\prime\prime}=\Phi(\beta^{\prime\prime})$, where, for each $i$, $\beta
_{i}^{\prime}=\pi_{i}^{\prime}-2$ and $\beta_{i}^{\prime\prime}=\pi
_{i}^{\prime\prime}-\max(2,\ell)$ such that $\ell$ is the maximal letter of $%
\pi^{\prime}$ (see Figure~\ref{Fphi1}(3)).

\item[\emph{(v)}] The contraction $\pi$ starts with two arcs, that is $%
\pi=12\pi^{\prime}1\pi^{\prime\prime}2\pi^{\prime\prime\prime}\in \mathcal{W}%
_{12}(n)$. We define $\Phi(\pi)$ to be the joint of the steps $LD$, the path
$P^{\prime}=\Phi(\beta^{\prime})$, the steps $LD$, the path $%
P^{\prime\prime}=\Phi(\beta^{\prime\prime})$, the steps $HDDHDD$, and the
path $P^{\prime\prime\prime}=\Phi(\beta^{\prime\prime\prime})$; for each $i$%
, $\beta_{i}^{\prime}=\pi_{i}^{\prime}-2$, $\beta_{i}^{\prime\prime}=\pi
_{i}^{\prime\prime}-\max(2,\ell^{\prime})$ and $\beta^{\prime\prime\prime}=%
\pi_{i}^{\prime\prime\prime}-max(2,\ell^{\prime},\ell^{\prime\prime})$, such
that $\ell^{\prime}$ and $\ell^{\prime\prime}$ are the maximal letters of $%
\pi^{\prime}$ and $\pi^{\prime\prime}$, respectively (see Figure~\ref{Fphi1}%
(4)).
\end{itemize}

The inverse map of $\Phi $ is clearly understood from the above cases (see
the factorizations of the paths in $\mathcal{P}_{n}$ and the factorizations
of the contractions in $\mathcal{W}_{12}(n)$ as described in Figure~\ref%
{Fphi1}).

\begin{theorem}
The map $\Phi$ is a bijection between the set of contractions in $\mathcal{W}%
_{12}(n)$ and the set of lattices paths in $\mathcal{P}_{n}$. Moreover, for
any contraction $\pi\in\mathcal{W}_{12}(n)$, we have:

\begin{itemize}
\item[\emph{(i)}] the number of arcs in the linear representation of $\pi$
equals the number of $HDD$ in the corresponding path $\Phi(\pi)$;

\item[\emph{(ii)}] the number of white vertices in the linear representation
of $\pi$ equals the number of $HD$ in the corresponding path $\Phi(\pi)$.
\end{itemize}
\end{theorem}

Let $P(x,y)=\sum_{n\geq 0}\sum_{P\in \mathcal{P}_{n}}x^{n}y^{HDD(P)}$ be the
generating function for the number of paths $P$ of length $3n$ in $\mathcal{P%
}_{n}$, respect to the semilength $n$ of $P$ and the number of occurrences $%
HDD(P)$ in the string $HDD$ in $P$. Then by the factorization of the paths
in $\mathcal{P}$, see Figure~\ref{Fphi1}, we obtain that the generating
function $P(x,y)$ satisfies
\begin{equation*}
P(x,y)=1+2xP(x,y)+2x^{2}yP^{2}(x,y)+x^{3}y^{2}P^{3}(x,y).
\end{equation*}%
From \eqref{eqrrf} we obtain that $P(x,y)=A_{1,s}(x,y,1)$. Thus we can state
the following result.

\begin{corollary}
The noncrossing normally ordered form of $(a+(a^{\dagger})^{2})^{n}$ is
given by
\begin{equation*}
\mathcal{NC}\biggl((a+(a^{\dag})^2)^{n}\biggr)=\sum_{j=0}^{n}\sum_{i=j}^{n}%
\frac{1}{n+1}\binom{n+1}{j+1}\binom{n-j}{n-i}\binom{2n-2i}{j}(a^{\dag
})^{2n-2i-j}a^{i-j}.
\end{equation*}
\end{corollary}

\subsection{$(a+a^{\dag})^{n}$, $(aa^{\dag})$, and 2-Motzkin paths}

A \emph{2-Motzkin path} of length $n$ is a path on the plane from the origin
$(0,0)$ to $(n,0)$ consisting

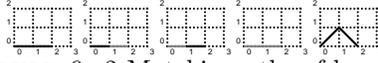
\begin{figure}[ht]
\begin{center}
\begin{pspicture}(0,0)(.9,.4)
\psgrid[unit=0.25,subgriddiv=1,griddots=5,gridlabels=3pt](0,0)(3,2)
\psline(0,0)(.5,0)
\end{pspicture}
\begin{pspicture}(0,0)(.9,.4)
\psgrid[unit=0.25,subgriddiv=1,griddots=5,gridlabels=3pt](0,0)(3,2)
\psline(0,0)(.25,0) \psline[linecolor=gray](.25,0)(.5,0)
\end{pspicture}
\begin{pspicture}(0,0)(.9,.4)
\psgrid[unit=0.25,subgriddiv=1,griddots=5,gridlabels=3pt](0,0)(3,2)
\psline(.25,0)(.5,0) \psline[linecolor=gray](0,0)(.25,0)
\end{pspicture}
\begin{pspicture}(0,0)(.9,.4)
\psgrid[unit=0.25,subgriddiv=1,griddots=5,gridlabels=3pt](0,0)(3,2)
\psline[linecolor=gray](0,0)(.5,0)
\end{pspicture}
\begin{pspicture}(0,0)(.9,.4)
\psgrid[unit=0.25,subgriddiv=1,griddots=5,gridlabels=3pt](0,0)(3,2)
\psline(0,0)(.25,.25)(.5,0)
\end{pspicture}
\end{center}
\par
\vspace{-14pt}
\caption{2-Motzkin paths of length $2$.}
\label{Mot1}
\end{figure}
of up steps, down steps, level steps colored black, and level steps colored
gray, such that the path does not go below the $x$-axis. We will use $U$, $D$%
, $L$, and $L^{\prime}$, to represent the up, down, black level, gray level
steps, respectively (see Figure~\ref{Mot1}). Most probably, the notion of
2-Motzkin path firstly appeared in the work of Delest and Viennot \cite{VD6}
and has been studied in a number of works, including \cite{VD1,VD8}. 


First, we describe a bijection $\Psi$ between the set of linear
representations $\mathcal{V}_{11}(n)$ and the set $\mathcal{M}_{n}$ of
2-Motzkin paths of length $n$. Let $\pi=\pi_{2n}\pi_{2n-1}\ldots\pi_{1}$ be
a linear representation in $\mathcal{V}_{11}(n)$. We read each time two
vertices from $\pi$ from right to left and successively generate the
2-Motzkin path. When a black vertex $\pi_{2j-1}$ with degree zero is
followed by a white vertex $\pi_{2j}$ with degree zero, then in the
2-Motzkin path we add a black level step $L$. When a black vertex $%
\pi_{2j-1} $ with degree one is followed by a white vertex $\pi_{2j}$ with
degree one, then in the 2-Motzkin path we add a gray level step $L^{\prime}$%
. When a black vertex $\pi_{2j-1}$ with degree one (resp. zero) is followed
by a white vertex $\pi_{2j}$ with degree zero (resp. one), then in the
2-Motzkin path we add an up (resp. down) step $U$ (resp. $D$). The reverse
of the map $\Psi$ is obvious.

\begin{proposition}
There is a bijection $\Psi$ between the set of noncrossing contractions in $%
\mathcal{V}_{11}(n)$ and the set of 2-Motzkin paths of length $n$. Moreover,
$\pi\in\mathcal{V}_{11}(n)$ has exactly $m$ edges if and only if the number
of black level steps and up-steps in $\Psi(\pi)$ equals $m$.
\end{proposition}

The above proposition gives that the noncrossing normally ordered form of $%
(aa^{\dag})^{n}$ is given by
\begin{equation*}
\mathcal{NC}\biggl((aa^{\dag})^{n}\biggr)=\sum\limits_{j=0}^{n}\frac{1}{n+1}%
\binom{n+1}{j+1}\binom{n+1}{j}(a^{\dag})^{n-j}a^{n-j},
\end{equation*}
(see Theorem~\ref{th33}) where the numbers $\frac{1}{n+1}\binom{n+1}{j+1}%
\binom{n+1}{j}$ are the so-called Narayana numbers (see, \emph{e.g.}, \cite%
{Z}).

Now, we describe a second bijection $\Theta$ between the set of linear
representations $\mathcal{W}_{11}(n)$ and the set $\mathcal{M}_{n}$ of
2-Motzkin paths of length $n$. Let $\pi=\pi_{2n}\pi_{2n-1}\ldots\pi_{1}$ be
a linear representation in $\mathcal{W}_{11}(n)$. We read $\pi$ from right
to left to generate the 2-Motzkin path. When a black (resp. white) vertex $%
\pi_{j}$ with degree zero is read, then in the 2-Motzkin path we add a black
(resp. gray) level step $L$ (resp. $L^{\prime}$). When a black (resp. white)
vertex with degree one is read, then in the 2-Motzkin path we add an up
(resp. down) step $U$ (resp. $D$). The reverse of the map $\Theta$ is again
obvious.

\begin{proposition}
There is a bijection $\Theta$, between the set of noncrossing contractions
in $\mathcal{W}_{11}(n)$ and the set of 2-Motzkin paths of length $n$.
Moreover, $\pi\in\mathcal{W}_{11}(n)$ has exactly $m$ edges and $d$ white
vertices if and only if the number of up-steps in $\Theta(\pi)$ equals $m$
and the number of gray level steps in $\Theta(\pi)$ is $d-m$.
\end{proposition}

The above proposition gives that the noncrossing normally ordered form of $%
(a+a^{\dag})^{n}$ is given by
\begin{equation*}
\mathcal{NC}\biggl((a+a^{\dag})^{n}\biggr)=\sum_{j=0}^{n}%
\sum_{i=j}^{n-j}c_{j}\binom{n}{i+j}\binom{i+j}{2j}(a^{\dag})^{n-i-j}a^{i-j}.
\end{equation*}
as described in Theorem \ref{thaa8}.

\section{Open problems}

A number of directions for further research seem to arise naturally. The first and foremost problem consists in deriving a more physical understanding of the process of noncrossing normal ordering. As discussed in Section~\ref{Prelim}, the result of noncrossing normal ordering cannot be reproduced by the conventional normal ordering where some kind of commutation relations is assumed. Thus, the statistics resulting from the noncrossing normal ordering is a new and nontrivial phenomenon which clearly deserves closer study. In this context it is not even clear whether one should call the operators for which the noncrossing normal ordering is applied ``bosonic'' anymore.

Let us also state two mathematical problems:

\begin{itemize}
\item Study the \emph{nonnesting normally ordered form} of a given
expression $F(a,a^{\dag })$. We say that the two edges $e=(a,b)$ and $%
e^{\prime }=(c,d)$ are \emph{nesting} if $a<c<d<b$ (that is, the edge $e$
covers the edge $e^{\prime }$) or $c<a<b<d$ (that is, the edge $e^{\prime }$
covers the edge $e$). In the case of nonnesting partitions, see~%
\cite{Kl1}.

\item Study the distribution of a given statistic on the set of normally
ordered form of a given expression. For example, study the asymptotic
behavior of the number of edges that cover other edges in the normally
ordered form of $(a(a^{\dag })^{r})^{n}$, when $n$ tends to infinity.
\end{itemize}

\bigskip

\noindent \emph{Acknowledgments.} The authors would like to thank Ed
Corrigan, Chris Fewster, Gherardo Piacitelli, and Tony Sudbery for
helpful discussion; Pawe\l\ B\l asiak and Jacob Katriel for
encouragement and interest in this work.


\end{document}